\documentclass[12pt]{article}
\usepackage{amsmath, amssymb}
\usepackage{mathbbol}
  \usepackage{epic,eepic}

\def\stareq{\buildrel\ast\over =}

\renewcommand{\a}{\alpha}
\renewcommand{\b}{\beta}
\newcommand{\g}{\gamma}
\newcommand{\s}{\sigma}
\newcommand{\ol}{\overline}
\newcommand{\bfsigma}{\mbox{\boldmath{$\sigma$}}}
\newcommand{\bfalpha}{\mbox{\boldmath{$\alpha$}}}
\newcommand{\bfgamma}{\mbox{\boldmath{$\gamma$}}}
\newcommand{\bfSigma}{\mbox{\boldmath{$\Sigma$}}}
\newcommand{\bfell}{\mbox{\boldmath{$\ell$}}}

\begin{document}

\title{The Gordon decompositions of the inertial currents of
  the Dirac electron correspond to a Foldy-Wouthuysen transformation}

\author{Ingo Kirsch\thanks{email: ik@thp.uni-koeln.de}\\Inst.\ Theor.\ 
 Physics, University of Cologne\\50923 K\"oln, Germany\\ \\ Lewis H.\ 
 Ryder\thanks{email: l.h.ryder@ukc.ac.uk} \and Friedrich W.\ 
 Hehl\thanks{Permanent address: Inst.\ Theor.\ Physics, University of
 Cologne, 50923 K\"oln, Germany., email: hehl@thp.uni-koeln.de}\\ 
 School of Physical Sciences,University of Kent\\ Canterbury CT2 7NR, UK}

\date{16 February 2001}
\maketitle

\newpage
\begin{abstract} We consider Dirac's free electron theory on the 
  first quantized level. We decompose its canonical spin current \`a
  la Gordon and find a {\em conserved} ``Gordon spin'' current which
  turns out to be equivalent to the Hilgevoord-Wouthuysen spin.  We
  can conclude therefrom that the Gordon-type decomposition mentioned
  above corresponds to a Foldy-Wouthuysen transformation which
  transforms the Dirac wave function from the conventional Dirac-Pauli
  to the Newton-Wigner representation.  {\em file lewis18.tex,
    2001-02-16}
\end{abstract}

\section{Introduction}
The question of the existence of a {\em conserved relativistic spin
  operator} for the free Dirac electron was answered positively for
the first time by Hilgevoord and Wouthuysen \cite{HilgevoordW}.  Their
spin operator turned out to be the relativistic generalization of the
non-relativistic Foldy-Wouthuysen spin operator \cite{FW}. They found
it by starting from the Lagrange formalism of the Klein-Gordon\mbox{
  (!)} field while treating the Dirac equation as a subsidiary
condition. This derivation has not been considered as completely
satisfactory.

Rather there have been demands for a group theoretical motivation for
a conserved spin operator.  The foundations for that were laid by
G\"ursey \cite{G2}.  The starting-point for such a derivation should
be the Pauli-Luba\'nski vector operator since its square is the spin
Casimir operator of the Poincar\'e group. Recently these ideas were
taken up again by Ryder \cite{LewisGRG} who defined a relativistic
operator which, in essence, equals the Foldy-Wouthuysen spin when
applied to positive energy states. In Section \ref{SecRyder}, we will
generalize the latter to a four dimensional covariant spin operator.
Unfortunately, it turns out that, in general, it is not conserved.

However, another method for obtaining a relativistic conserved spin
density, from which the corresponding operator can easily be deduced,
has been known for quite some time \cite{Seitz}: By the method of
Gordon \cite{Gordon}, we decompose the energy-momentum and spin
current densities of the Dirac electron into convective and
polarization parts.  Moreover, we find the translational and Lorentz
{\em gravitational moments} of the Dirac electron.  We can show that
the convective spin current, Gordon spin for short, is conserved.
Therefore, we expect it to be equivalent to the Hilgevoord-Wouthuysen
spin. We will prove that in Section \ref{SecHW}.

The paper is organized in the following way. Section \ref{Sec2} is
devoted to the classical field theory of the currents of the Dirac
electron including the Gordon-type decompositions of its inertial
currents.  It concludes with the proof of the equivalence of the
Gordon spin and the Hilgevoord-Wouthuysen spin.  Section \ref{Sec3}
recalls the Foldy-Wouthuysen transformation of the Dirac wave function
which leads to the conserved Foldy-Wouthuysen mean spin operator.
Section 4 covers its relativistic extension, the Hilgevoord-Wouthuysen
spin, and shows its relation to the G\"ursey-Ryder spin. In Section
\ref{Sec4}, we conclude that the Gordon decompositions of the
energy-momentum and spin currents of the Dirac electron correspond to
a Foldy-Wouthuysen transformation of its wave function.

\subsection*{Notation}
  
We work in {\em flat} Minkowski space, i.e.\ without gravity, and
consider a {\em free} Dirac electron. 

The conventions with respect to the Dirac matrices and their
representation are taken from Ryder \cite{Lewisbook}, in particular,
$\sigma^{ij}=\frac{i}{2}[\gamma^i,\gamma^j]$. Commutator and
anticommutator read $[A,B]=AB-BA$ and $\{A,B\}=AB+BA$, respectively.
Conventions: Minkowski metric $\eta^{\a\b}=(+--\,-)$, orthonormal
frame $e_\alpha$ with
$\alpha,\beta,\dots=\hat{0},\hat{1},\hat{2},\hat{3}$ (frame indices),
components $e^i{}_\alpha$, here $i,j,\dots=0,1,2,3$ are coordinate
indices. Coframe $\vartheta^\beta$ with components $e_j{}^\beta$.
Spatial coordinates $a,b,\dots=1,2,3$. $c=\hbar=1$. Parenthesis around
indices denote symmetrization $(ij)=\frac{1}{2}(ij+ji)$, brackets
antisymmetrization $[ij]=\frac{1}{2}(ij-ji)$.  The covariant
derivative is $D_\a=e^i{}_\a D_i$ with $D_i=\partial_i
+\frac{i}{4}\,\Gamma_{i\a\beta}\,\sigma^{\a\beta}$, the
$\Gamma_{i\a\b}$ are orthonormal frame components of the (flat)
Riemannian connection, with $\Gamma_{i\a\b}=-\Gamma_{i\b\a}$. A ``star
equal'' $\stackrel{*}{=}$ denotes equality the validity of which is
restricted to a certain coordinate system or frame.

\section{Currents within classical field theory} \label{Sec2}

\subsection{Noether currents of energy-momentum \& spin \cite{Corson}}

We start with a matter field $\Psi$ and its Lagrangian density
\begin{equation}\label{Dlagrangian}
  {\mathcal L} ={\mathcal L}\left( \eta_{\alpha\beta},\vartheta^\alpha,\Psi,
    \ol\Psi, D_\alpha\Psi, D_\alpha\ol\Psi\right)\,.
\end{equation}
The corresponding action
\begin{equation}\label{action}
  W=\int {\mathcal L} \, d^4x
\end{equation}
is invariant under Poincar\'e transformations, i.e., in particular
under translations and Lorentz transformations. Then, by Noether's
theorem, we have the conservation of energy-momentum 
\begin{equation}\label{emcons}
  D_i\,\Sigma_\alpha{}^i = 0\,
\end{equation}
and of angular momentum\footnote{We follow here the conventions of
  Ryder \cite{Lewisbook}. In ref.\ \cite{Schuecking}, the conventions
  are such that the orbital angular momentum reads $x_{[\a}
  \Sigma_{\b]}$.}
\begin{equation}\label{amcons}
  D_i\left(\tau_{\alpha\beta}{}^i+x_\alpha\Sigma_\beta{}^i-
    x_\beta\Sigma_\alpha{}^i \right) \stackrel{*}{=} 0\,.
\end{equation}
Here the conserved total angular momentum density of the field $\Psi$
splits into a {\em spin} part, $\tau_{\a\b}{}^i$, and an {\em orbital}
part, $2 x_{[\a} \Sigma_{\beta]}{}^i$.  

These four plus six conservation laws hold ``weakly'', i.e.\ if the
matter field equation $\delta {\mathcal L}/ \delta \Psi=0$ is satisfied.
The canonical energy-momentum and spin (angular momentum) currents are
defined by
\begin{equation}\label{canem}
  \Sigma_\alpha{}^i:= \delta_\alpha^i{\mathcal L}-\frac{\partial{\mathcal L}}{
    \partial D_i\Psi}\,D_\alpha\Psi-D_\alpha\overline{\Psi} \,
  \frac{\partial{\mathcal L}}{\partial D_i\overline{\Psi}}
\end{equation} and
\begin{equation}\label{canam}
  \tau_{\alpha\beta}{}^i := -\frac{\partial{\mathcal L}}{\partial D_i
    \Psi}\, \ell_{\alpha\beta}\Psi + \overline{\Psi}\,
  \ell_{\alpha\beta}\, \frac{\partial{\mathcal L}}{\partial D_i
    \overline{\Psi}}\,,
\end{equation} 
respectively, where $\ell_{\alpha\beta}:=\frac{i}{2}\s_{\a\b}
=-\ell_{\beta\alpha}$ are the generators of the Lorentz group.

The energy-momentum current $\Sigma_\alpha{}^i$ has 16 independent
components in general. We lower the index $i$ and find $\Sigma_{\a\b}
=e_{i\beta} \Sigma_{\a}{}^i$. It can be decomposed irreducibly under
the Lorentz group into a {\em trace} $\Sigma=\Sigma_{\g}{}^\g$, a {\em
  traceless symmetric} $\Sigma_{(\a\b)}- \frac{1}{4}\eta_{\a\b}
\Sigma$ and an {\em antisymmetric} piece $\Sigma_{[\a\b]}$. In a
similar way the spin density with its 24 independent components
decomposes into three pieces \cite{Seitz},
\begin{equation}
  \tau_{\alpha\beta\gamma}=^{(1)}\!\tau_{\alpha\beta\gamma}+
  ^{(2)}\!\tau_{\alpha\beta\gamma}+^{(3)}\!\tau_{\alpha\beta\gamma}
  \,,
\end{equation}
where
\begin{align}
  ^{(1)}\tau_{\alpha\beta\gamma}&=\tau_{\alpha\beta\gamma}-\frac{2}{3}
  \tau_{[\alpha\vert\delta}{}^{\delta}\,\eta_{\vert\beta]\gamma}-
  \tau_{[\alpha\beta\gamma]}\quad&&{\rm (16\,\,components)} \,,\\ 
  ^{(2)}\tau_{\alpha\beta\gamma}&=\frac{2}{3}
  \tau_{[\alpha\vert\delta}{}^{\delta}\,\eta_{\vert\beta]\gamma}\quad&&{\rm
    (4\,\,components)} \,,\\ 
  ^{(3)}\tau_{\alpha\beta\gamma}&=\tau_{[\alpha\beta\gamma]}\quad&&{\rm
    (4\,\,components)} \,.
\end{align}

In eq.(\ref{amcons}) we differentiate the second and third term.  We
use $D_i x^\a \stackrel{*}{=} \delta_i^\a$ and apply (\ref{emcons}).
Then we find a simpler form of the angular momentum conservation law,
\begin{equation}\label{conssymm}
  D_i\,\tau_{\alpha\beta}{}^i -2 \Sigma_{[\alpha\beta]}=0\,.
\end{equation}
Thus, if the energy-momentum tensor is symmetric, i.e.\ $
\Sigma_{[\alpha\beta]}=0$, then already the spin current
$\tau_{\alpha\beta}{}^i$ is separately conserved --- and not only the
total angular momentum current, see (\ref{amcons}).

\subsection{Relocalization of energy-momentum \& spin and
superpotentials}

The Noether currents $\Sigma_\alpha{}^i$ and $\tau_{\alpha\beta}{}^i$
are only determined up to gradients. If we add gradients to
$\Sigma_\alpha{}^i$ and $\tau_{\alpha\beta}{}^i$, we call it a {\em
  relocalization} of momentum and spin. We have the following lemmas,
see ref.\cite{ROMP}:

\subsubsection*{Lemma 1:} The canonical or Noether
currents fulfill the conservation laws
\begin{equation}\label{cons10}
  D_i\,\Sigma_\alpha{}^i=0\,,\qquad
  D_i\,\tau_{\alpha\beta}{}^i- 2 \Sigma_{[\alpha\beta]}=0\,.
\end{equation}
The {\em relocalized} currents
\begin{eqnarray}
\hat\Sigma_\alpha{}^i( X ) &=&\Sigma_\alpha{}^i-D_jX
  _\alpha{}^{ij}\,,\label{reloc1}\\ \hat\tau_{\alpha\beta}{}^i( X ,Y)
  &=&\tau_{\alpha\beta}{}^i-2 X _{[\alpha\beta]}{}^i
  -D_jY_{\alpha\beta}{}^{ij}\,,\label{reloc2}
\end{eqnarray}
satisfy the analogous conservation laws
\begin{equation}\label{cons'}
  D_i\,\hat\Sigma_\alpha{}^i=0\,,\qquad
  D_i\,\hat\tau_{\alpha\beta}{}^i-2 \hat
  \Sigma_{[\alpha\beta]}=0\,.
\end{equation}
The superpotentials $X_\alpha{}^{ij}(x)=-X_\alpha{}^{ji}$ and
$Y_{\alpha\beta}{}^{ij}(x) =-Y_{\beta\alpha}{}^{ij}
=-Y_{\alpha\beta}{}^{ji}$ represent $24+36$ {\em arbitrary} functions.

\subsubsection*{Lemma 2:} The total energy-momentum 
\begin{equation}
  P_\a:\stackrel{*}{=}\int_{H_t} \Sigma_\a{}^i\, dS_i
\end{equation}
and the total angular momentum 
\begin{equation}
  J_{\alpha\beta}:\stackrel{*}{=}\int_{H_t} (\tau_{\alpha\beta}{}^i+
  2 x_{[\alpha}\Sigma_{\beta]}{}^i)\,dS_i
\end{equation}
are invariant under relocalization,
\begin{equation}
  \hat P_\alpha\stareq P_\alpha-\int_{\partial H_t} X _\alpha{}^{ij}\,
  da_{ij}\qquad \hat J_{\alpha\beta}\stareq
  J_{\alpha\beta}-\int_{\partial H_t} \left(2 x_{[\alpha}\, X
    _{\beta]}{}^{ij}+Y_{\alpha\beta}{}^{ij} \right)da_{ij}\,,
\end{equation}
provided the superpotentials $X_\alpha{}^{ij}$ and
$Y_{\alpha\beta}{}^{ij}$ approach zero at spacelike asymptotic
infinity sufficiently fast. Here $H_t$ denotes a spacelike
hypersurface in Minkowski space with 3-volume element $dS_i$ and
$\partial H_t$ its 2-dimensional boundary with area element
$da_{ij}=-da_{ji}$. Orthonormal frames are used throughout.

\subsubsection*{Belinfante currents as example}

The most straightforward approach to a relocalization is to put both,
$\hat\tau_{\alpha\beta}{}^i=0$ and $Y_{\alpha\beta}{}^{ij}=0$. Then we
find the Belinfante currents
\begin{equation}\label{}
  t_\alpha{}^i:=\;\stackrel{{\rm B}}{\Sigma}_\alpha{}^{\!i}
  (\stackrel{{\rm B}}{X}) \qquad{\rm and}\qquad\stackrel{{\rm
      B}}{\tau}_{\alpha\beta}{}^i =0= {\tau}_{\alpha\beta}{}^{\!i} -
  2 \stackrel{{\rm B}}{X}_{[\alpha\beta]}{}^{\!i}
\end{equation}
or
\begin{equation}\label{}
  \stackrel{{\rm B}}{X}_\alpha{}^{\!ij}= \frac{1}{2} \left( \tau_\alpha{}^{ij}-
  \tau^{ij}{}_\alpha + \tau^j{}_\alpha{}^i \right) \,.
\end{equation}
We collect our results with respect to the Belinfante relocalization in
\begin{equation}\label{belinfante}
  D_i\,t_\alpha{}^i=0\,,\quad t_{[\alpha\beta]}=0\,,\quad t_\alpha{}^i
  = \Sigma_\alpha{}^i-\frac{1}{2} D_j \left(\tau_\alpha{}^{ij}-
    \tau^{ij}{}_\alpha + \tau^j{}_\alpha{}^i \right) \,.
\end{equation}
This relocalization can be understood as one which kills the
Belinfante spin current, i.e., the relocalized total angular momentum
under this particular conditions reduces to its orbital part alone.

\subsection{The inertial currents of the Dirac electron}

The explicit form of the Dirac Lagrangian density reads \cite{Lewisbook} 
\begin{equation}\label{Dlagrangian'}
  {\mathcal L}_{\rm D} = \frac{i}{2}\left[\overline{\Psi}\gamma^\alpha
    (D_\alpha\Psi)- (D_\alpha\overline{\Psi})\gamma^\alpha\Psi
  \right]-m\,\overline{\Psi}\Psi\,.
\end{equation}
We substitute this expression into (\ref{canem}) and (\ref{canam}) and
use also the Dirac equation. Then we find
\begin{equation}\label{Dcanem}
  \Sigma_\alpha{}^i= \frac{i}{2} \left[(D_\alpha\overline{\Psi})
    \gamma^i\Psi - \overline{\Psi} \gamma^i D_\alpha\Psi \right]\, ,
\end{equation} 
\begin{equation}\label{Dcanspin}
  \tau_{\alpha\beta}{}^i=\frac{1}{4}\,\overline{\Psi} ( \gamma^i
  \sigma_{\alpha\beta} + \sigma_{\alpha\beta} \gamma^i ) \Psi \,.
\end{equation}
The canonical spin current can be alternatively written as
\begin{equation}\label{}
  \tau_{\alpha\beta\gamma}=\tau_{[\alpha\beta\gamma]}=
  -\frac{1}{2}\,\epsilon_{\alpha\beta\gamma \delta}
  \overline{\Psi}\gamma_5\gamma^\delta\Psi\,
  =\frac{i}{2}\,\overline{\Psi}\gamma_{[\a}\gamma_{\b}\gamma_{\g]}\Psi.
\end{equation}
We recognize thereby that the canonical spin depends on only 4
independent components. The reason for this is that the electron is a
fundamental particle.  Such particles are described by irreducible
parts of tensors which are fundamental in a mathematical sense, in
this case the axial part $^{(3)}\tau_{\alpha\beta\gamma}$.

The dual of the spin current $\tau_{\alpha\beta\gamma}$ is the {\em axial} 
spin current 
\begin{align}
  S^{\delta}=\frac{1}{3!} \varepsilon^{\alpha\beta\gamma\delta}
  \tau_{\alpha\beta\gamma}=\frac{1}{2} \overline{\Psi} \gamma_5
  \gamma^\delta \Psi \,.
\end{align}
When we split $\Psi$ into two 2-spinors $\psi$ and $\chi$, it reduces to
the non-relativistic Pauli spin density, see \cite{Lewisbook}, p.55,
\begin{equation}\label{Dspin}
 {S}_a = \frac{1}{2}\psi^\dagger \sigma_a\psi\,,
\end{equation}
with $\sigma_a$, $a=1,2,3$ as the Pauli matrices.

Since the canonical spin is totally antisymmetric, it follows
immediately from (\ref{belinfante}b,c) that the Belinfante current
$t_\alpha{}^i$ for a Dirac electron is the symmetric part of the
canonical current $\Sigma_\a{}^i$,
\begin{equation}
  t_{\a\b}=\Sigma_{(\a\b)}\,,
\end{equation}
with $t_{\a\b}=e_{i\b}\, t_\a{}^i$ and $\Sigma_{\a\b}= e_{i\b}\,
\Sigma_\a{}^i$.  

The inertial currents, i.e.\ the densities of the Dirac momentum and
the Dirac spin $(\Sigma_\alpha{}^i,\tau_{\alpha\beta}{}^i)$, enter as
sources on the right hand sides of the field equations of the {\em
  Einstein-Cartan theory} of gravity \cite{Erice1}. In other words,
spin, besides mass-energy, is a source of gravity. In general
relativity, however, mass-energy is the only source of gravity.  In
this case, one has to take the (symmetric) Belinfante energy-momentum
current as the source term in general relativity, thereby excluding
spin from being gravitationally active.  Both theories, the
Einstein-Cartan theory and general relativity, are viable since they
are indistinguishable by current observation.

\subsection{Gordon decomposition (relocalization) of the currents of
the Dirac field}

Gordon \cite{Gordon} decomposed the Dirac current
$\overline{\Psi}\gamma^i \Psi$ into a {\em convective} and a {\em
  polarization} part. An analogous procedure was applied, first by
Markov \cite{Markov}, to the energy-momentum current of the Dirac
field. Later, in the Erice lectures \cite{Erice1}, the Gordon
decompositions of the energy-momentum {\em and} the spin currents of
the Dirac electron were displayed, following earlier results of von
der Heyde \cite{vdHeyde}. Therefrom, for $m\neq0$, the translational
and the Lorentz gravitational moment densities were extracted as
\begin{equation}\label{TrMom}
  M_\alpha{}^{ij}= \frac{i}{4m}\, \left[
    \overline{\Psi}\sigma^{ij}D_\alpha\Psi-D_\alpha \overline{\Psi}
    \sigma^{ij}\Psi \right]
\end{equation}
and
\begin{equation}\label{LorMom}
  M_{\alpha\beta}{}^{ij} = \frac{1}{8m}\, \overline{\Psi} (\sigma^{ij}
  \sigma_{\alpha\beta} +\sigma_{\alpha\beta} \sigma^{ij})\Psi \,,
\end{equation} respectively. Seitz \cite{Seitz} redid these calculations
and showed that this is a universally valid procedure which can also
be applied to the energy-momentum and spin currents of fields with
spin $1,\frac{3}{2}$, and $2$; for further developments see, e.g.,
\cite{Schuecking,Lemke,Obukhov1}.

The gravitational moment densities (\ref{TrMom}, \ref{LorMom}), in the
sense of the relocalization of Lemma 1, correspond to the choices
\begin{equation}\label{supgor1}
  X _\alpha{}^{ij} = M_\alpha{}^{ij}+2 \delta_\alpha^{[i}\,M_k^{\cdot
    j]k}
\end{equation}
and
\begin{equation}\label{supgor2}
  Y_{\alpha\beta}{}^{ij}=M_{\alpha\beta}{}^{ij}\,.
\end{equation}
Accordingly, for the relocalized currents we eventually find
\begin{equation}\label{sigmag}
  \stackrel{{\rm G}}{\Sigma}_\alpha{}^{\!i} =
  \delta_\alpha^i\stackrel{{\rm G}}{{\mathcal L}}-
  \frac{1}{2m}\left(D^i\overline{\Psi} D_\alpha\Psi +
    D_\alpha\overline{\Psi} D^i \Psi \right)
\end{equation}
and
\begin{equation}\label{gordonspin}
  \stackrel{{\rm G}}{\tau}_{\alpha\beta}{}^{\!i} = \frac{1}{4mi}
  \left[ \left(D^i\overline{\Psi}\right) \sigma_{\alpha\beta} \Psi -
    \overline{\Psi} \sigma_{\alpha\beta} \left( D^i \Psi \right)
  \right] \,.
\end{equation}
Here we have the Gordon Lagrangian            
\begin{equation}\label{}
  \stackrel{{\rm G}}{{\mathcal
      L}}\,:=\frac{1}{2m}\left[(D_\alpha\overline{\Psi}
    )D^\alpha\Psi-m^2\overline{\Psi}\Psi \right]\,.
\end{equation}

In summary, for the Gordon type momentum and spin currents, we have
\begin{equation}\label{}
  D_i\stackrel{{\rm G}}{\Sigma}_\alpha{}^{\!i}= 0\,,\qquad
  \stackrel{{\rm G}}{\Sigma}_{[\alpha\beta]}= 0\,,\qquad
  D_i\stackrel{{\rm G}}{\tau}_{\alpha\beta}{}^{\!i}= 0\,.
\end{equation}
This seems to be the only way one can derive a relativistic {\em spin}
density which is automatically {\em conserved} by itself. The
Belinfante momentum $t_\alpha{}^i$ is also symmetric, similarly to the
Gordon momentum $\stackrel{{\rm G}}{\Sigma}_\alpha{}^{\!i}$. Therefore
spin is conserved. However, in the Belinfante case the relocalized
spin vanishes, i.e., the statement is trivial. Not so in the Gordon
case: Here we have a non-vanishing conserved spin derived by means of
a procedure within the framework of standard Lagrangian field theory.

It is not clear to us which type of gravitational theory is induced if
the Gordon currents $(\stackrel{{\rm G}}{\Sigma}_\alpha{}^{\!i},
\stackrel{{\rm G}}{\tau}_{\alpha\beta}{}^{\!i})$ are the sources of
gravity.  Perhaps it is only the Einstein-Cartan theory rewritten in a
suitable way.

\subsection{Gordon spin $\longrightarrow$ Hilgevoord-Wouthuysen spin}
\label{SecHW}

\subsubsection*{Hermitian HW-spin}
Now that we have a conserved spin, we recall that Hilgevoord and
Wouthuysen \cite{HilgevoordW} had already derived the conserved spin
``charge''
\begin{equation}\label{HWop}
  \stackrel{{\rm HW}}{S^{\alpha\beta}}=\frac{1}{4} \int \ol{\Psi}
  \left\{ \gamma^0 ,\sigma^{\a\b} \right\} \Psi d^3x + \frac{i}{4m}
  \int \left( D_a \ol{\Psi} \gamma^a \gamma^0 \s^{\a\b} \Psi -
    \ol{\Psi} \s^{\a\b} \gamma^0 \gamma^a D_a \Psi \right) d^3x\,,
\end{equation}
here written in its hermitian form, see \cite{HilgevoordW} eq.(3.1).
The corresponding spin {\em density} reads
\begin{eqnarray} \label{HWspinherm}
  \stackrel{{\rm HW}}{S^{\alpha\beta i}}:=\frac{1}{4}
  \overline\Psi\{\gamma^i,\sigma^{\alpha\beta}\}\Psi\!\!\!&+&\!\!\!\frac{i}{4m}
  (D_j \overline\Psi \gamma^j\gamma^i\sigma^{\alpha\beta} \Psi-
  \overline\Psi \sigma^{\alpha\beta}\gamma^i \gamma^j D_j
  \Psi)\nonumber\\ 
  \!\!\!&-&\!\!\!\frac{i}{4m}(D^i\overline\Psi\sigma^{\alpha\beta}\Psi-
  \overline\Psi\sigma^{\alpha\beta} D^i\Psi) \,.
\end{eqnarray} 
Indeed, integration of (\ref{HWspinherm}) over a space-like
hypersurface,
\begin{align} \label{integration}
\int_{H_t} \stackrel{{\rm HW}}{S^{\alpha\beta i}} dS_i =
\int_{H_t} \stackrel{{\rm HW}}{S^{\alpha\beta 0}} dS_0 =
\,\stackrel{{\rm HW}}{S^{\alpha\beta}}\,,
\end{align}
leads back to the HW-spin (\ref{HWop}), since $dS_0=d^3 x$. Here we
inserted $\gamma^0 \gamma^0=\mathbb{1}$ in each of the two terms in
the second line of (\ref{HWspinherm}); then, 
$D^i\overline\Psi\sigma^{\alpha\beta}\Psi=D^i\overline\Psi\gamma^0
\gamma^0 \sigma^{\alpha\beta}\Psi$, e.g. Moreover, we used $\gamma^i
D_i=\gamma^0 D_0+\gamma^a D_a$.

Let us prove that the Gordon spin (\ref{gordonspin}) coincides with
(\ref{HWspinherm}).  We multiply the Dirac equation
\begin{equation} \label{Dirac}
  i \g^ j D_j \Psi= m \Psi
\end{equation}
and its adjoint 
\begin{equation}
  -i D_j \ol \Psi \g^j=m \ol \Psi \, 
\end{equation}
by $-i \g^i$ from the left and  $i \g^i$ from the right,
respectively. Using \cite{Thaller}, eq.(2.220), namely
\begin{equation} \label{Th2220}
  \g^i \g^j = \eta^{ij} \mathbb{1} - i \s^{ij} \, ,
\end{equation}
we obtain
\begin{eqnarray}
  D^i \Psi &=& -i m \g^i \Psi + i \s^{ij} D_j \Psi \\ D^i \ol \Psi &=&
  i m \ol \Psi \g^i + i D_j \ol \Psi \s^{ji} \, .
\end{eqnarray}
By substituting this into (\ref{gordonspin}), we find
\begin{eqnarray}\label{tautauG}
  \stackrel{{\rm G}}{\tau}{}^{\!\!\a\b i}&=&\frac{1}{4mi} \left[(im
    \ol\Psi \g^i + i D_j \ol \Psi \s^{ji}) \s^{\a\b} \Psi - \ol \Psi
    \s^{\a\b} (-i m \g^i \Psi + i \s^{ij} D_j \Psi)\right] \nonumber\\ 
  &=&\frac{1}{4} \ol \Psi \{ \s^{\a\b}, \g^i \} \Psi + \frac{1}{4m} [
  D_j \ol \Psi \s^{ji} \s^{\a\b} \Psi - \ol \Psi \s^{\a\b} \s^{ij} D_j
  \Psi ] \, .
\end{eqnarray}
The $\s^{ij}$ in the second term are replaced by means of (\ref{Th2220}).
This yields 
\begin{eqnarray}\label{GordonHW}
  \stackrel{{\rm G}}{\tau}{}^{\!\!\a\b i}=\frac{1}{4} \overline\Psi
  \{\gamma^i,\sigma^{\alpha\beta}\}\Psi\!\!\!&+&\!\!\!\frac{i}{4m}
  (D_j \overline\Psi \gamma^j\gamma^i\sigma^{\alpha\beta} \Psi-
  \overline\Psi \sigma^{\alpha\beta}\gamma^i \gamma^j D_j
  \Psi)\\ 
  \!\!\!&-&\!\!\!\frac{i}{4m}(D_j\overline\Psi\eta^{ij}
  \sigma^{\alpha\beta}\Psi- \overline\Psi\sigma^{\alpha\beta}\eta^{ij}
  D_j\Psi) =\,\stackrel{{\rm HW}}{S^{\alpha\beta i}} \,, \nonumber
\end{eqnarray}
i.e.\ the Gordon spin current really coincides with the hermitian
HW-spin.

\subsubsection*{Non-hermitian HW-spin}
Since the Gordon spin (\ref{gordonspin}) is hermitian, we have first 
considered the hermitian form of the HW-spin charge. Shorter and
more common is the non-hermitian HW-spin charge
\begin{equation}\label{HdK}
  \stackrel{{\rm HW}}{S^{\alpha\beta}}=\int\Psi^\dagger
  \left[\frac{1}{2}\,\sigma^{\alpha\beta}+
    \frac{1}{2m}\left(\gamma^\alpha D^\beta-\gamma^\beta D^\alpha
    \right)\right]\Psi\,d^3x\,,
\end{equation}
see \cite{HilgevoordW}, eq.(3.2), and also \cite{HilgevoorddK}. The
corresponding spin density reads
\begin{equation} \label{HWspin}
  \stackrel{{\rm HW}}{S^{\alpha\beta i}_{\rm nh}}:=\ol
  \Psi \g^i \left(\frac{1}{2}  \s^{\a\b}+\frac{1}{m} \g^{[\a} D^{\b]}
  \right) \Psi\,.
\end{equation} 
Again, by means of an integral of the type (\ref{integration}) and
$\ol\Psi\gamma^0= \Psi^\dagger \gamma^0 \gamma^0 = \Psi^\dagger$, the
HW-spin charge (\ref{HdK}) can be deduced therefrom.

What is the relation of the Gordon spin (\ref{gordonspin}) to
(\ref{HWspin})? Partial integration of (\ref{tautauG}) yields
\begin{eqnarray}
  \stackrel{{\rm G}}{\tau}{}^{\!\!\a\b i} &=&\frac{1}{4} \ol \Psi \{
  \s^{\a\b}, \g^i \} \Psi - \frac{1}{4m} \left( - \ol \Psi \s^{ij}
    \s^{\a\b} D_j \Psi + \ol\Psi \s^{\a\b}\s^{ij} D_j \Psi \right) -
  D_j \tilde Y^{\a\b ij} \nonumber\\ &=&\frac{1}{4} \ol \Psi (2 \g^i
  \s^{\a\b}+ [\s^{\a\b},\g^i])\Psi - \frac{1}{4m}
  \left(\ol\Psi[\s^{\a\b},\s^{ij}] D_j \Psi\right) - D_j \tilde
  Y^{\a\b ij}
    \label{HWG} \, ,
\end{eqnarray} 
with
\begin{eqnarray}
  \tilde Y^{\a\b ij}:= \frac{1}{4m} (\ol \Psi \s^{ij} \s^{\a\b} \Psi).
\end{eqnarray} 
By means of the identity
\begin{equation} \label{identity}
  \frac{1}{4} \ol \Psi([\s^{\a\b},\g^i] )\Psi-\frac{1}{4m} (\ol \Psi
  [\s^{\a\b}, \s^{ij}] D_j \Psi) = \frac{1}{2m}\ol \Psi (\g^i \g^\a
  D^\b - \g^i \g^\b D^\a) \Psi\,,
\end{equation} 
see Appendix \ref{Appendix1}, we finally get
\begin{eqnarray} 
  \stackrel{{\rm G}}{\tau}{}^{\!\!\a\b i} &=&\frac{1}{2} \ol \Psi [
  \g^i \s^{\a\b} + \frac{1}{m} (\g^i \g^\a D^\b - \g^i \g^\b D^\a) ]
  \Psi - D_j \tilde Y^{\a\b ij} \nonumber\\ &=&\stackrel{{\rm
      HW}}{S^{\alpha\beta i}_{\rm nh}} -\, D_j \tilde Y^{\a\b ij}\,.
\end{eqnarray}
Result: We have shown that the Gordon spin (\ref{gordonspin}) is the
same as the hermitian Hilgevoord-Wouthuysen spin (\ref{HWspinherm}).
We also have proven its equivalence to the non-hermitian HW-spin
density (\ref{HWspin}) up to a total divergence. In an integral of the
total spin, the divergence vanishes by means of the Gauss theorem.

\section{Unitarily transforming the Dirac wave function} \label{Sec3}
In the last section we discussed the {\em currents} of the Dirac field
and deduced the gravitational moments by means of the Gordon
decomposition. 
We found that there is a kind of ``gauge'' freedom in defining
these currents.  The Gordon decomposition made use of this freedom and
led finally to a separately conserved spin current.
We now turn to a different description of the Dirac field. Instead
of studying the currents, we address the Dirac wave function
itself. In this context we deal with operators and
their expectation values, i.e.\ quantities of the form
\begin{eqnarray}
O=\int \Psi^\dagger (x)\, O_{\rm op}\, \Psi(x)\, d^3 x \,.
\end{eqnarray}
In the Dirac theory there is another arbitrariness as far as the
representation of the wave function is concerned. One representation,
by means of a {\em canonical transformation}, can be transformed to a
second one leaving the physical quantities invariant. As we will see
below, there is a representation of the Dirac wave equation in which
the spin operator $\frac{1}{2} \bfsigma$ is a constant of motion
separately.

\subsection{Dirac-Pauli representation}
In relativistic quantum mechanics we start usually with the 
  Dirac-Pauli representation of the Dirac equation.  The latter is
either given by its well-known covariant form
\begin{equation}\label{}
\left( i \gamma^\a D_\a - m \right) \Psi = 0
\end{equation}
or by its Schr\"odinger form with the Hamiltonian $H_{\rm DP}$:
\begin{equation}\label{}
i \frac{\partial \Psi}{\partial t}=H_{\rm DP} \Psi\,,\quad
H_{\rm DP}:= \beta m + {\bf \bfalpha}\cdot{\bf p}.
\end{equation}
It is the only representation of the Dirac equation which is linear in
the momentum {\bf p}. The advantage is that the {\em minimal coupling}
of the electromagnetic field to the Dirac field, in this
representation, agrees with experiment. A priori we have no way of
knowing which representation the minimal coupling scheme applies to,
since minimal coupling is a representation dependent scheme.
The reason is that, in general, a {\em canonical transformation} is
momentum dependent. Therefore, canonical transformations and minimal
coupling are noncommuting operations, i.e., they lead to different
wave equations, see Costella and McKellar \cite{Costella}.

The disadvantage of the Dirac-Pauli representation is that neither a
``decent'' position operator nor a ``decent'' velocity operator can be
defined. A position measurement, based on the usual position operator
{\bf x}, would lead to pair production, if carried out below the
Compton wavelength, since the Dirac equation couples positive and
negative energy states. The creation of particles and antiparticles
hinders an exact position measurement. Also the velocity operator does
not make any classical sense, since its eigenvalues are plus or minus
the speed of light. In contrast, the velocity of a classical Dirac
particle lies in between these extremal values.

\subsection{Newton-Wigner representation}
These deficiencies were cured by Newton and Wigner \cite{NW}.
Originally, the aim of these authors was to formulate the properties
of {\em localized states} for particles with arbitrary spin on the
basis of invariance requirements. They write \cite{NW}:
\begin{quote}
{\it Chief of these is that a state, localized at a certain
point, becomes, after a translation, orthogonal to all the undisplaced
states localized at that point. It is found that the required
properties uniquely define the set of localized states for elementary
systems of non-zero mass and arbitrary spin.}
\end{quote}
Indeed, they have shown that such states can be found uniquely for
arbitrary spin and that these states have all purely positive (or,
equivalently, purely negative) energy. These states form the
continuous eigenvalue spectrum of a particular operator
\begin{quote}
{\em ... which it is natural to call the position operator. This
  operator has automatically the property of preserving the positive
  energy character of the wave function to which it is applied (and it
  should be applied only to such wave functions).}
\end{quote}
Thus, in contrast to the usual position operator {\bf x} in the
Dirac-Pauli representation, this operator, which we denote by {\bf X},
preserves the definiteness of the energy since no
particle-antiparticle pairs are created. In addition, the
eigenvalues of the corresponding velocity operator ${\bf V}:=d{\bf
  X}/dt$ range from minus to plus the speed of light. Due to these
properties, the operator {\bf X} is a position operator in the
classical sense.

Thus, for a Dirac particle, it would be of interest to find the
canonical transformation leading to the representation of the Dirac
equation in which the operator {\bf x} becomes the Newton-Wigner
position operator {\bf X}.

\subsection{Foldy-Wouthuysen transformation and mean spin}
This was implemented by Foldy and Wouthuysen \cite{FW}.  They found
the canonical (unitary) transformation of the wave function,
\begin{eqnarray}\label{}
  \Psi'&=&e^{iS} \Psi, \\ H'&=&e^{iS} \left(H-i
    \frac{\partial}{\partial t}\right) e^{-iS},
\end{eqnarray}
where $S$ is a hermitian operator. It transforms the Dirac-Pauli
representation of the free Dirac Hamiltonian,
\begin{equation}\label{}
  H_{\rm DP}= \beta m + {\bf \bfalpha}\cdot{\bf p},
\end{equation}
into its Newton-Wigner representation\footnote{Also called
  Foldy-Wouthuysen representation. These considerations were
  generalized by Obukhov \cite{Obukhov2} in the case of the 
  existence of an external gravitational field.},
\begin{equation}\label{}
  H_{\rm NW}= \beta \sqrt{m^2 + {\bf p}^2} .
\end{equation}
In contrast to the Dirac-Pauli representation, the Newton-Wigner
representation contains only the {\em even} operator $\beta$.  Even
operators do not couple the upper with the lower two components of a
Dirac spinor (in the standard representation). Therefore the
components for positive and negative energy states are completely
decoupled and the position operator in the Newton-Wigner
representation preserves the sign of the energy.

Any operator in the Dirac-Pauli representation is related to that in
the Newton-Wigner representation by
\begin{equation}\label{}
  O_{\rm DP}= e^{-iS} \, O_{\rm NW} \, e^{iS} \, .
\end{equation}
Thus the corresponding position operator in the Dirac-Pauli
representation reads, see \cite{FW}, eq.(23),
\begin{equation}\label{}
  {\bf X}= e^{-iS} \,{\bf x}\, e^{iS} = {\bf x} + \textmd{oscillating
    part}.
\end{equation}
It is the one also found by Newton and Wigner. It is called {\em
  mean-position} operator, since the original position operator {\bf
  x} consists of two parts: the mean-position operator {\bf X} and a
part oscillating rapidly about zero with an amplitude of about the
Compton wavelength (``Zitterbewegung'').

Among other interesting properties, the position operator {\bf X} can
be used for defining a new angular momentum operator. Remember
that the orbital angular momentum operator ${\bfell}={\bf x} \times
{\bf p}$ and the spin operator
\begin{equation}
  \frac{1}{2} \bfsigma \times \mathbb{1}=\frac{1}{2}\begin{pmatrix}
    \bfsigma &0\\0&\bfsigma\end{pmatrix}=\frac{1}{2i} \bfalpha \times
  \bfalpha, \quad \textmd{Pauli matrices }\bfsigma\, ,
\end{equation}
in the Dirac-Pauli representation, are no constants of
motion separately. However, the operators
\begin{equation}
  {\bf L}=e^{-iS} \,{\bfell}\, e^{iS}= {\bf X} \times {\bf p}
\end{equation}
and\footnote{In the following, we will write $\bfsigma$ instead of
  $\bfsigma \times \mathbb{1}$.}
\begin{equation} \label{FWNW}
  \bfSigma= e^{-iS} \frac{1}{2}\,{\bfsigma}\, e^{iS}= \frac{1}{2}
  \bfsigma - \frac{i (\bfgamma \times {\bf p})}{2E} -\frac{{\bf p}
    \times (\bfsigma \times {\bf p})}{2E(E+m)}
\end{equation}
are separately conserved in time, since they commute with the Dirac
Hamiltonian $H_{\rm DP}$. The operator $\bfSigma$ is called {\em mean
  spin operator}\footnote{Redefined by a factor $\frac{1}{2}$ because
  of different conventions.} and can also be written as (use
$\bfsigma=\gamma_5 \gamma^0 \bfgamma$),
\begin{equation}\label{FWop}
  \bfSigma=\frac{m}{2E} \bfsigma+ \frac{\bfsigma \cdot {\bf p}}
  {2E(E+m)} {\bf p}+\frac{i}{2E} \gamma_5 \gamma^0 \bfsigma \times
  {\bf p}.
\end{equation}
In the rest frame, in which ${\bf p}=0$ and $E=m$, this operator
reduces to $\frac{1}{2} {\bfsigma}$.

To sum up, the spin operator $\frac{1}{2} {\bfsigma}$ shall only be
used in the Newton-Wigner representation, since here it is conserved,
$[H_{\rm NW},\frac{1}{2} {\bfsigma}]=0$.  In the Dirac-Pauli
representation, however, we have to use the Foldy-Wouthuysen spin
\mbox{operator $\bfSigma$}, since $[H_{\rm DP},\bfSigma]=0$.

\section{Different spin operators}
\subsection{Hilgevoord-Wouthuysen spin}
Subsequent to the work of Foldy and Wouthuysen, it was Hilgevoord and
Wouthuysen \cite{HilgevoordW} who searched for a conserved covariant
spin in relativistic quantum field theory\footnote{For earlier work on
  relativistic conserved spin tensors, see Fradkin and Good
  \cite{Fradkin2} and references given therein.}. The expectation
value of the spin operator, the spin ``charge'', which involves an
integration over a three-dimensional space-like hypersurface, must
transform in a covariant manner. This is only guaranteed if the spin
current is conserved. As shown above, cf.\ eq.(\ref{conssymm}), this
is the case if the energy-momentum current is symmetric. However, the
canonical energy-momentum current of the Dirac electron is not
symmetric, see eq.(\ref{Dcanem}). Thus the canonical spin current
(\ref{Dcanspin}) is not conserved and a different splitting of the
total angular momentum in an orbital and a spin part is needed in such
a way that both parts are conserved.

Remember that every solution of the Dirac equation is also
a solution of the Klein-Gordon equation
\begin{align}
  (D^\alpha D_\alpha+m^2) \Psi=0 \,.
\end{align}
Therefore, one can start with the Klein-Gordon Lagrangian
\begin{equation}\label{}
  {\mathcal L}=\frac{1}{2m}\left[(D_\alpha\overline{\Psi}
    )D^\alpha\Psi-m^2\overline{\Psi}\Psi \right]\,
\end{equation}
and treat the Dirac equation as a subsidiary condition on the
solutions. It turns out that the canonical energy-momentum and spin
currents of the Klein-Gordon Lagrangian are the same as the Gordon
currents (\ref{sigmag}) and (\ref{gordonspin}).  The energy-momentum
is symmetric and thus spin is separately conserved.  The spin operator
can then be defined as 
\begin{equation}\label{spinHW}
  \stackrel{{\rm HW}}{S^{\alpha\beta}}=\int \tau^{\alpha\beta0}_{\rm
    KG} \,d^3x\,=\frac{1}{4mi} \int \left[
    \left(D^0\overline{\Psi}\right) \sigma^{\alpha\beta} \Psi -
    \overline{\Psi} \sigma^{\alpha\beta} \left( D^0 \Psi \right)
  \right] \,d^3x\,.
\end{equation}
The time derivatives can be eliminated by using the Dirac equation
in the form
\begin{equation}
  i \gamma^0 D_0 \Psi=(-i \gamma^a D_a + m) \Psi \,.
\end{equation}
Then (\ref{spinHW}) becomes
\begin{equation}
  \stackrel{{\rm HW}}{S^{\alpha\beta}}=\frac{1}{4} \int \ol{\Psi}
  \left\{ \gamma^0 ,\sigma^{\a\b} \right\}
  \Psi d^3x - \frac{i}{4m} \int \left( \ol{\Psi} \s^{\a\b} \gamma^0
      \gamma^a D_a \Psi - D_a \ol{\Psi} \gamma^a \gamma^0 \s^{\a\b}
      \Psi \right) d^3x\,,
\end{equation}
see \cite{HilgevoordW} eqs.(3.1) and (3.2), or, equivalently,
\begin{equation}\label{HWspinop}
  \stackrel{{\rm HW}}{S^{\alpha\beta}}=\int \Psi^\dagger
  S^{\alpha\beta}_{\rm op} \Psi d^3x, \quad S^{\alpha\beta}_{\rm op}=
  \frac{1}{2}\,\sigma^{\alpha\beta}+ \frac{i}{2m}\left(p^\alpha
    \gamma^\beta-p^\beta \gamma^\alpha \right) \, .
\end{equation}
The spatial part of this operator is given by $S_a=\frac{1}{2}
\varepsilon_{abc} S_{bc}^{\rm op}$ or, explicitly, by
\begin{equation}\label{HW3D}
  {\bf S} = \frac{1}{2} \bfsigma + \frac{i}{2m} {\bf p} \times
  \bfgamma \,.
\end{equation}

Note that it is not necessary to require the operator
$S^{\alpha\beta}_{\rm op} $ to be covariant. Indeed,
$S^{\alpha\beta}_{\rm op} $ is not a Lorentz tensor, since we 
have
\begin{eqnarray} 
  \left[S^{\rm op}_{\alpha\beta}, \frac{i}{2} \s_{\gamma\delta} \right]&=&
    \eta_{\alpha\gamma}\,S^{\rm op}_{\beta\delta} -
    \eta_{\alpha\delta}\,S^{\rm op}_{\beta\gamma} +
    \eta_{\beta\delta}\,S^{\rm op}_{\alpha\gamma}
    -\eta_{\beta\gamma}\,S^{\rm op}_{\alpha\delta} \nonumber\\
  &&+ \frac{i}{m} (\eta_{[\a|\g} p_{\delta} \g_{|\b]}
                  -\eta_{[\a|\delta} p_{\g} \g_{|\b]} )\,.
\end{eqnarray}
Covariance and conservation of the corresponding spin {\em density}
are sufficient.   


\subsection{G\"ursey-Ryder spin operator}\label{SecRyder}
As first pointed out by Wigner \cite{Wigner}, the most satisfactory
relativistic definition of spin is that it generates the ``little
group'' of the Poincar\'e (or inhomogeneous Lorentz) group; this is
the group that leaves a given 4-momentum invariant.  Wigner showed, on
general grounds, that the little group for timelike momenta is
$SU(2)$, but he did not find expressions for the operators which
generate the algebra of this group.  To find these operators, it is
sensible to start from the Pauli-Luba\'nski operator
\begin{equation}\label{PauliL}
  W_\alpha:=\frac{1}{2}\,\varepsilon_{\alpha\beta\gamma\delta}J^{\beta\gamma}
  P^\delta\,,
\end{equation}
since this operator involves the ten generators $J_{\a\b}$ and
$P_\gamma$ of the Poincar\'e group. We then define the following
tensor operators and their duals
\begin{equation}\label{W}
  W_{\alpha\beta}:=\left[W_\alpha,\, W_\beta \right]\,,\qquad
  W^{\ast}_{\alpha\beta}:=\frac{1}{2}\,\varepsilon_{\alpha\beta\gamma\delta}
  W^{\gamma\delta}\,,
\end{equation}
with
\begin{equation}\label{W*}
  W_{\alpha\beta}^{\ast\ast}=-W_{\alpha\beta}\,.
\end{equation}
We are now looking for operators which generate (for timelike momenta)
the group $SU(2)$, and therefore satisfy the commutation relations
\begin{equation}\label{CR}
\left[X_a,X_b\right]=i\epsilon_{abc}X_c\,.
\end{equation}
The four operators $W_\a$ clearly do not generate this algebra. Let us
define the operators
\begin{equation}\label{XY}
  X_{\alpha\beta}:=-\frac{i}{m^2}
  \left(W_{\alpha\beta}+iW^{\ast}_{\alpha\beta} \right)\,,\qquad
  Y_{\alpha\beta}:=-\frac{i}{m^2}
  \left(W_{\alpha\beta}-iW^{\ast}_{\alpha\beta} \right)\,,
\end{equation}
which are, up to the constant factor $-\frac{i}{m^2}$, the
{\em anti-selfdual and selfdual} parts of $W_{\a\b}$, i.e.\ they satisfy
\begin{equation}\label{duality}
  X_{\alpha\beta}^{\ast}=-iX_{\alpha\beta}\,,\qquad
  Y_{\alpha\beta}^\ast= iY_{\alpha\beta}\,.
\end{equation}
It can be checked, and was first pointed out by G\"ursey \cite{G2},
that these operators $X_{\a\b}$ and $Y_{\a\b}$ {\em both} generate the
algebra of $SO(1,3)$
\begin{equation}\label{CR1}
  \left[X_{\alpha\beta},X_{\gamma\delta} \right]=-i\left(
    \eta_{\alpha\gamma}\,X_{\beta\delta} -
    \eta_{\alpha\delta}\,X_{\beta\gamma} +
    \eta_{\beta\delta}\,X_{\alpha\gamma}
    -\eta_{\beta\gamma}\,X_{\alpha\delta} \right)\,.
\end{equation}
It then follows that $X_a = \frac{1}{2} \varepsilon_{abc} X_{bc}$ and
$Y_a = \frac{1}{2} \varepsilon_{abc} Y_{bc}$ both obey the commutation
relations of $SU(2)$, eqn.\ (\ref{CR}) above.  We now take the step of
identifying these as the spin operators for the left- and right-handed
parts of a Dirac spinor.  To this end, we define the operator
\begin{equation}\label{Gue4D}
  Z_{\a\b}:=\frac{1}{2}(1-\gamma_5) X_{\a\b} + \frac{1}{2}(1+\gamma_5)
  Y_{\a\b}\,.
\end{equation}
It clearly obeys the commutation relations of $SO(1,3)$, eqn.\ 
(\ref{CR1}). And, following the same logic, the three operators
\begin{equation}
  Z_a:=\frac{1}{2}(1-\gamma_5) X_a + \frac{1}{2} (1+\gamma_5) Y_a
\end{equation}
will generate $SU(2)$, so $\mathbf{Z}$ is therefore a plausible
candidate for a spin operator for Dirac particles.  In fact it was
shown in \cite{LewisGRG} that, {\em acting on positive energy states},
$\mathbf{Z}$ is the same as the Foldy-Wouthuysen mean spin operator;
see remark c) and d) of section \ref{sec43}. In particular it follows
that $\mathbf{Z}$ is conserved, since the FW spin operator is
conserved.  When not acting on $E > 0$ states, however, $Z_{\a\b}$ is
not conserved.  In fact it may be shown that
\begin{align} \label{relationZ&S}
  Z_{\a\b}=2 S^{\rm op}_{\a\b} - \frac{1}{2} \sigma_{\a\b} \,,
\end{align}
see appendix eq.(\ref{relZ&S}), where $S^{\rm op}_{\a\b}$ is given by
(\ref{HWspinop}). The HW-spin $S^{\rm op}_{\a\b}$ is conserved, but
$\sigma_{\a\b}$ is not. Consequently, $Z_{\a\b}$, although it has
the desirable property of generating the $SO(1,3)$ algebra, {\em is not a
conserved quantity}. $S^{\rm op}_{\a\b}$ on the other hand is
conserved, but does not generate $SO(1,3)$; in fact it may be shown
that
\begin{equation} \label{HWcommutator}
  \left[S^{\rm op}_{\alpha\beta},S^{\rm op}_{\gamma\delta} \right]=-i\left(
    \eta_{\alpha\gamma}\,S^{\rm op}_{\beta\delta} -
    \eta_{\alpha\delta}\,S^{\rm op}_{\beta\gamma} +
    \eta_{\beta\delta}\,S^{\rm op}_{\alpha\gamma}
    -\eta_{\beta\gamma}\,S^{\rm op}_{\alpha\delta} \right) +
  \frac{1}{m^2} [ p_{[\alpha} \gamma_{\beta]}, p_{[\gamma}
  \gamma_{\delta]}] \,.
\end{equation}
It is only when the operator $Z_a$ acts on positive energy states that
we are able to find an operator that is both conserved and obeys the
$SU(2)$ commutation relations - and is derived explicitly from
considerations of Lorentz covariance.

\subsection{Comparing different spin operators} \label{sec43}
\begin{table}
\caption{Various spin operators.} \label{table}
\begin{center}
\begin{tabular}{|l|l|l|} 
\hline
\bf operator & \bf perpendicular & \bf parallel\\
\hline
  \raisebox{-1.4ex}{Foldy-Wouthuysen $\bfSigma$}
& \begin{minipage}{4cm} 
   $$ \bfSigma _\perp = \frac{m-\bf{p} \cdot \bfgamma}{E}
   \frac{1}{2} \bfsigma_\perp $$ 
  \end{minipage}
& \begin{minipage}{2cm}
  $$\bfSigma _\parallel=\frac{1}{2} \bfsigma_\parallel$$
  \end{minipage} \\
 \raisebox{-1.4ex}{Hilgevoord-Wouthuysen ${\bf S}$} 
& \begin{minipage}{4cm}
  $${\bf S_\perp}=\frac{E}{m} \frac{m- {\bf p} \cdot \bfgamma}{E}
  \frac{1}{2} \bfsigma_\perp $$
  \end{minipage}
& \begin{minipage}{2cm}
  $${\bf S_\parallel}=\frac{1}{2} \bfsigma_\parallel$$
  \end{minipage} \\
 \raisebox{-1.4ex}{polarization operator $\bf O$} 
& \begin{minipage}{4cm}
  $${\bf O_\perp}=\gamma^0 \frac{1}{2} \bfsigma_\perp$$
  \end{minipage} 
& \begin{minipage}{2cm}
  $${\bf O_\parallel}=\pm \frac{1}{2} \bfsigma_\parallel$$
   \end{minipage} \\
  \raisebox{-1.4ex}{G\"ursey-Ryder $\bf Z$}  
& \begin{minipage}{4cm} 
  $${\bf Z_\perp}=\frac{ E-\gamma^0 \bfgamma \cdot \bf p}{m} 
  \frac{1}{2} \bfsigma_\perp \smallskip$$
  \end{minipage}  
& \begin{minipage}{2cm}
  $${\bf Z_\parallel}=\frac{1}{2} \bfsigma_\parallel \smallskip $$
  \end{minipage}\\
\hline
\end{tabular}
\end{center}
\end{table}
The Foldy-Wouthuysen spin $\bfSigma$, eq.(\ref{FWNW}), the
Hilgevoord-Wouthuysen spin ${\bf S}$, eq.(\ref{HW3D}), the G\"ursey-Ryder
spin $\bf Z$, given explicitly by \cite{LewisGRG} eq.(21),
\begin{equation} \label{Guer3D}
{\bf Z}=\frac{E}{2m} \bfsigma - \frac{{\bf p}(\bfsigma \cdot 
{\bf p})}{2m(E+m)}  - i \gamma_5 \frac{\bfsigma \times {\bf p}}{2m}\,,
\end{equation}
and the polarization operator\footnote{Redefined by a
  factor $\frac{1}{2}$ due to different conventions.} ${\bf O}$
\cite{StechI,Fradkin1,Fradkin2}
\begin{equation}\label{Polarization}
{\bf O}=\gamma^0 \frac{\bfsigma}{2} - \frac{\gamma_5}{2E} {\bf p} - \gamma^0
\frac{{\bf p}}{2E(E+m)} (\bfsigma \cdot {\bf p})\,,
\end{equation}
are all separately conserved spin operators, though constructed in
different ways. Recall that $\bf Z$ is only conserved when acting on
positive energy states. Hence, we expect close relationships between
these operators.  To see them, we divide each of the operators in
parts perpendicular and parallel to the \mbox{momentum ${\bf p}$}, see
\cite{FW}.  We summarize the results in Tab.\ \ref{table}.

\begin{figure}
\begin{center}
  \setlength{\unitlength}{0.00083333in}
\begingroup\makeatletter\ifx\SetFigFont\undefined%
\gdef\SetFigFont#1#2#3#4#5{%
  \reset@font\fontsize{#1}{#2pt}%
  \fontfamily{#3}\fontseries{#4}\fontshape{#5}%
  \selectfont}%
\fi\endgroup%
{\renewcommand{\dashlinestretch}{30}
\begin{picture}(6013,3489)(0,-10)
\path(375,987)(375,687)
\blacken\path(570.000,492.000)(450.000,462.000)(570.000,432.000)(570.000,492.000)
\path(450,462)(1425,462)
\blacken\path(1305.000,432.000)(1425.000,462.000)(1305.000,492.000)(1305.000,432.000)
\path(3750,2937)(3750,2637)
\path(3750,2037)(3750,1737)
\path(3750,987)(3750,687)
\blacken\path(2895.000,492.000)(2775.000,462.000)(2895.000,432.000)(2895.000,492.000)
\path(2775,462)(3600,462)
\blacken\path(3480.000,432.000)(3600.000,462.000)(3480.000,492.000)(3480.000,432.000)
\blacken\path(1845.000,492.000)(1725.000,462.000)(1845.000,432.000)(1845.000,492.000)
\path(1725,462)(2400,462)
\blacken\path(2280.000,432.000)(2400.000,462.000)(2280.000,492.000)(2280.000,432.000)
\blacken\path(1695.000,1317.000)(1575.000,1287.000)(1695.000,1257.000)(1695.000,1317.000)
\path(1575,1287)(2925,1287)
\blacken\path(2805.000,1257.000)(2925.000,1287.000)(2805.000,1317.000)(2805.000,1257.000)
\blacken\path(1695.000,3192.000)(1575.000,3162.000)(1695.000,3132.000)(1695.000,3192.000)
\path(1575,3162)(2925,3162)
\blacken\path(2805.000,3132.000)(2925.000,3162.000)(2805.000,3192.000)(2805.000,3132.000)
\path(4725,1887)(4800,1887)(4800,12)(4725,12)
\path(4725,2487)(4800,2487)(4800,1962)(4725,1962)
\path(4725,3462)(4800,3462)(4800,2562)(4725,2562)
\put(675,462){\makebox(0,0)[lb]{\smash{{{\SetFigFont{12}{14.4}{\rmdefault}{\mddefault}{\updefault}pos.\ en.}}}}}
\put(750,312){\makebox(0,0)[lb]{\smash{{{\SetFigFont{12}{14.4}{\rmdefault}{\mddefault}{\updefault}states}}}}}
\put(3000,462){\makebox(0,0)[lb]{\smash{{{\SetFigFont{12}{14.4}{\rmdefault}{\mddefault}{\updefault}boost}}}}}
\put(4875,237){\makebox(0,0)[lb]{\smash{{{\SetFigFont{12}{14.4}{\rmdefault}{\mddefault}{\updefault}(pos.\ energy)}}}}}
\put(4875,462){\makebox(0,0)[lb]{\smash{{{\SetFigFont{12}{14.4}{\rmdefault}{\mddefault}{\updefault}3-operators}}}}}
\put(4875,1287){\makebox(0,0)[lb]{\smash{{{\SetFigFont{12}{14.4}{\rmdefault}{\mddefault}{\updefault}relativistic}}}}}
\put(4875,1062){\makebox(0,0)[lb]{\smash{{{\SetFigFont{12}{14.4}{\rmdefault}{\mddefault}{\updefault}operators}}}}}
\put(4875,1512){\makebox(0,0)[lb]{\smash{{{\SetFigFont{12}{14.4}{\rmdefault}{\mddefault}{\updefault}covariant}}}}}
\put(4875,2112){\makebox(0,0)[lb]{\smash{{{\SetFigFont{12}{14.4}{\rmdefault}{\mddefault}{\updefault}charges}}}}}
\put(4875,2787){\makebox(0,0)[lb]{\smash{{{\SetFigFont{12}{14.4}{\rmdefault}{\mddefault}{\updefault}densities}}}}}
\put(4875,3012){\makebox(0,0)[lb]{\smash{{{\SetFigFont{12}{14.4}{\rmdefault}{\mddefault}{\updefault}current}}}}}
\put(0,3237){\makebox(0,0)[lb]{\smash{{{\SetFigFont{12}{14.4}{\rmdefault}{\mddefault}{\updefault}Gordon spin $\stackrel{{\rm G}}{\tau}_{\alpha\beta}{}^{\!i}$ }}}}}
\put(300,3012){\makebox(0,0)[lb]{\smash{{{\SetFigFont{12}{14.4}{\rmdefault}{\mddefault}{\updefault}eq.(\ref{gordonspin})}}}}}
\put(0,1362){\makebox(0,0)[lb]{\smash{{{\SetFigFont{12}{14.4}{\rmdefault}{\mddefault}{\updefault}G\"ursey-Ryder $Z_{\a\b}$}}}}}
\put(300,1137){\makebox(0,0)[lb]{\smash{{{\SetFigFont{12}{14.4}{\rmdefault}{\mddefault}{\updefault}eq.(\ref{Gue4D})}}}}}
\put(300,387){\makebox(0,0)[lb]{\smash{{{\SetFigFont{12}{14.4}{\rmdefault}{\mddefault}{\updefault}{\bf Z}}}}}}
\put(225,162){\makebox(0,0)[lb]{\smash{{{\SetFigFont{12}{14.4}{\rmdefault}{\mddefault}{\updefault}eq.(\ref{Guer3D})}}}}}
\put(1500,387){\makebox(0,0)[lb]{\smash{{{\SetFigFont{12}{14.4}{\rmdefault}{\mddefault}{\updefault}{\bf O}}}}}}
\put(1350,162){\makebox(0,0)[lb]{\smash{{{\SetFigFont{12}{14.4}{\rmdefault}{\mddefault}{\updefault}eq.(\ref{Polarization})}}}}}
\put(2250,612){\makebox(0,0)[lb]{\smash{{{\SetFigFont{12}{14.4}{\rmdefault}{\mddefault}{\updefault}Foldy-W. }}}}}
\put(2475,387){\makebox(0,0)[lb]{\smash{{{\SetFigFont{12}{14.4}{\rmdefault}{\mddefault}{\updefault}$\bfSigma$}}}}}
\put(2400,162){\makebox(0,0)[lb]{\smash{{{\SetFigFont{12}{14.4}{\rmdefault}{\mddefault}{\updefault}eq.(\ref{FWop})}}}}}
\put(3675,387){\makebox(0,0)[lb]{\smash{{{\SetFigFont{12}{14.4}{\rmdefault}{\mddefault}{\updefault}{\bf S}}}}}}
\put(3450,162){\makebox(0,0)[lb]{\smash{{{\SetFigFont{12}{14.4}{\rmdefault}{\mddefault}{\updefault}eq.(\ref{HW3D})}}}}}
\put(1950,312){\makebox(0,0)[lb]{\smash{{{\SetFigFont{12}{14.4}{\rmdefault}{\mddefault}{\updefault}$\gamma^0$}}}}}
\put(3075,1362){\makebox(0,0)[lb]{\smash{{{\SetFigFont{12}{14.4}{\rmdefault}{\mddefault}{\updefault}HW spin op. $S^{\a\b}_{\rm op}$}}}}}
\put(2075,1362){\makebox(0,0)[lb]{\smash{{{\SetFigFont{12}{14.4}{\rmdefault}{\mddefault}{\updefault}$\frac{1}{2} \sigma_{\a\b}$}}}}}
\put(3450,1137){\makebox(0,0)[lb]{\smash{{{\SetFigFont{12}{14.4}{\rmdefault}{\mddefault}{\updefault}eq.(\ref{HWspinop})}}}}}
\put(3150,3237){\makebox(0,0)[lb]{\smash{{{\SetFigFont{12}{14.4}{\rmdefault}{\mddefault}{\updefault}``HW'' spin $\stackrel{{\rm HW}}{S^{\alpha\beta i}}$}}}}}
\put(3450,3012){\makebox(0,0)[lb]{\smash{{{\SetFigFont{12}{14.4}{\rmdefault}{\mddefault}{\updefault}eq.(\ref{HWspinherm})}}}}}
\put(3150,2337){\makebox(0,0)[lb]{\smash{{{\SetFigFont{12}{14.4}{\rmdefault}{\mddefault}{\updefault}HW spin charge}}}}}
\put(3450,2112){\makebox(0,0)[lb]{\smash{{{\SetFigFont{12}{14.4}{\rmdefault}{\mddefault}{\updefault}eq.(\ref{HdK})}}}}}

\end{picture}
} \caption{The zoo of various spin densities and
    operators.}
\label{diagram}
\end{center}
\end{figure}
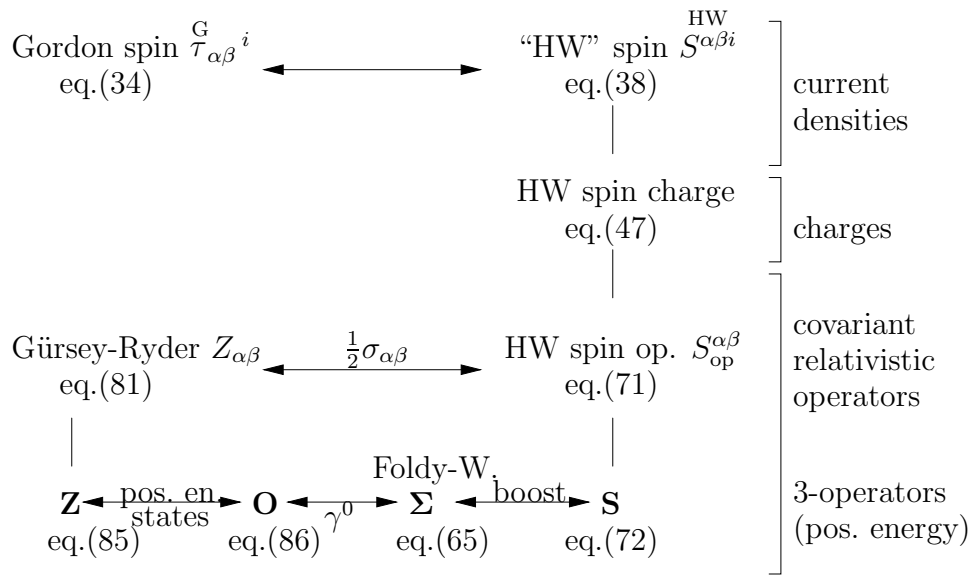

When we restrict the application of these three-operators 
to positive energy states, we make the following observations: 
\begin{itemize}
\item[a)] As observed first by G\"ursey \cite{G1}, the FW mean spin
  operator $\bfSigma$, which was defined as a nonrelativistic
  operator, is also relativistic in this case.
\item[b)] The above table shows that ${\bf S_\perp}$ and
  $\bfSigma_\perp$ just differ by the Lorentz factor
  \mbox{$\frac{E}{m}=\frac{1} {\sqrt{1-v^2}}$}, while the parallel
  parts are the same.  Thus the axial vector ${\bf S}$ is equal to
  ${\bfSigma}$, when boosted to the rest frame. Thus it is the
  laboratory-system operator for the spin of the particle in its own
  rest frame.
\item[c)] When applied to positive energy states, those for which
\begin{equation}
  \gamma^i p_i =\gamma^0 E - \bfgamma \cdot {\bf p}= m \,,
\end{equation}
the G\"ursey-Ryder spin operator $\bf Z$ becomes identical to the
polarization operator $\bf O$, see the table.
\item[d)] Moreover, in the Newton-Wigner representation both equal the
  Foldy-Wout\-huysen operator $\bfSigma$ up to the factor $\gamma^0$
  which is just $\pm 1$ for positive and negative energy states,
  respectively, i.e. $\bfSigma_{\rm NW}=\gamma^0 {\bf O}_{\rm
    NW}=\frac{1}{2}\bfsigma$, cf.\ (\ref{FWNW}) with ref.
  \cite{Fradkin1} eq.(7.3). Thus $\bfSigma={\bf O}={\bf Z}$ for
  positive energy.
\end{itemize}
\mbox{Fig.\ \ref{diagram}} summarizes all relations between the conserved 
spin densities and operators considered in this paper.

\section{Discussion}\label{Sec4}

Let us remind ourselves that Newton-Wigner \cite{NW} looked
successfully for a decent {\em position operator}. Foldy-Wouthuysen
\cite{FW} implemented this explicitly by transforming the Dirac-Pauli
representation of the electron into a new representation, now
appropriately called Newton-Wigner representation, in which the
position operator applied to electron states with positive energy
preserves the positive energy.

Since orbital angular momentum can only be defined in a reasonable way
when a decent position operator is available --- remember ${\bf
  L}={\bf X}\times{\bf p}$ --- an invariant splitting of the total
angular momentum can seemingly only be achieved in the Newton-Wigner
representation.  This led finally to the (nonrelativistic)
Foldy-Wouthuysen spin operator. Subsequently, Hilgevoord and
Wouthuysen derived a {\em conserved} spin operator which turned out to
be the relativistic extension of the Foldy-Wouthuysen spin operator.
It is remarkable that Hilgevoord-Wouthuysen used a technique which
strictly was outside of the Lagrange formalism.

The relocalization \`a la Gordon yields a spin density that is the
same as the one found by Hilgevoord-Wouthuysen, see (\ref{GordonHW}).
However, the spatial part of the Hilgevoord-Wouthuysen spin operator
is related to the Foldy-Wouthuysen spin operator just by a simple
Lorentz transformation.

\begin{figure}
\begin{center}
  \setlength{\unitlength}{0.00083333in}
\begingroup\makeatletter\ifx\SetFigFont\undefined%
\gdef\SetFigFont#1#2#3#4#5{%
  \reset@font\fontsize{#1}{#2pt}%
  \fontfamily{#3}\fontseries{#4}\fontshape{#5}%
  \selectfont}%
\fi\endgroup%
{\renewcommand{\dashlinestretch}{30}
\begin{picture}(5151,2088)(0,-10)
\put(4275,1704){\makebox(0,0)[lb]{\smash{{{\SetFigFont{12}{14.4}{\rmdefault}{\mddefault}{\updefault}densities}}}}}
\put(4275,1929){\makebox(0,0)[lb]{\smash{{{\SetFigFont{12}{14.4}{\rmdefault}{\mddefault}{\updefault}current}}}}}
\put(4275,1104){\makebox(0,0)[lb]{\smash{{{\SetFigFont{12}{14.4}{\rmdefault}{\mddefault}{\updefault}relativistic}}}}}
\put(4275,879){\makebox(0,0)[lb]{\smash{{{\SetFigFont{12}{14.4}{\rmdefault}{\mddefault}{\updefault}operators}}}}}
\put(4275,129){\makebox(0,0)[lb]{\smash{{{\SetFigFont{12}{14.4}{\rmdefault}{\mddefault}{\updefault}3-operators}}}}}
\path(1125,204)(2625,204)
\blacken\path(2505.000,174.000)(2625.000,204.000)(2505.000,234.000)(2505.000,174.000)
\put(1125,279){\makebox(0,0)[lb]{\smash{{{\SetFigFont{12}{14.4}{\rmdefault}{\mddefault}{\updefault}Foldy-Wouthuysen}}}}}
\put(1275,54){\makebox(0,0)[lb]{\smash{{{\SetFigFont{12}{14.4}{\rmdefault}{\mddefault}{\updefault}transformation}}}}}
\path(375,1629)(375,1329)
\path(375,804)(375,504)
\path(1125,1854)(2625,1854)
\blacken\path(2505.000,1824.000)(2625.000,1854.000)(2505.000,1884.000)(2505.000,1824.000)
\path(3300,1629)(3300,1329)
\path(3300,804)(3300,504)
\put(75,1704){\makebox(0,0)[lb]{\smash{{{\SetFigFont{12}{14.4}{\rmdefault}{\mddefault}{\updefault}eq.(\ref{Dcanspin})}}}}}
\put(75,279){\makebox(0,0)[lb]{\smash{{{\SetFigFont{12}{14.4}{\rmdefault}{\mddefault}{\updefault}Pauli spin}}}}}
\put(150,54){\makebox(0,0)[lb]{\smash{{{\SetFigFont{12}{14.4}{\rmdefault}{\mddefault}{\updefault}$\frac{1}{2} \bfsigma$}}}}}
\put(150,1029){\makebox(0,0)[lb]{\smash{{{\SetFigFont{12}{14.4}{\rmdefault}{\mddefault}{\updefault}$\frac{1}{2} \sigma_{\a\b}$}}}}}
\put(0,1929){\makebox(0,0)[lb]{\smash{{{\SetFigFont{12}{14.4}{\rmdefault}{\mddefault}{\updefault}can. spin $\tau_{\alpha\beta}{}^i$}}}}}
\put(1275,1704){\makebox(0,0)[lb]{\smash{{{\SetFigFont{12}{14.4}{\rmdefault}{\mddefault}{\updefault}decomposition}}}}}
\put(1500,1929){\makebox(0,0)[lb]{\smash{{{\SetFigFont{12}{14.4}{\rmdefault}{\mddefault}{\updefault}Gordon }}}}}
\put(2700,1929){\makebox(0,0)[lb]{\smash{{{\SetFigFont{12}{14.4}{\rmdefault}{\mddefault}{\updefault}Gordon spin $\stackrel{{\rm G}}{\tau}_{\alpha\beta}{}^{\!i}$}}}}}
\put(2700,1104){\makebox(0,0)[lb]{\smash{{{\SetFigFont{12}{14.4}{\rmdefault}{\mddefault}{\updefault}HW spin $S^{\a\b}_{\rm op}$}}}}}
\put(2925,879){\makebox(0,0)[lb]{\smash{{{\SetFigFont{12}{14.4}{\rmdefault}{\mddefault}{\updefault}eq.(\ref{HWspinop})}}}}}
\put(2925,1704){\makebox(0,0)[lb]{\smash{{{\SetFigFont{12}{14.4}{\rmdefault}{\mddefault}{\updefault}eq.(\ref{gordonspin})}}}}}
\put(2775,279){\makebox(0,0)[lb]{\smash{{{\SetFigFont{12}{14.4}{\rmdefault}{\mddefault}{\updefault}FW-spin $\bfSigma$ }}}}}
\put(2925,54){\makebox(0,0)[lb]{\smash{{{\SetFigFont{12}{14.4}{\rmdefault}{\mddefault}{\updefault}eq.(\ref{FWop})}}}}}
\end{picture}
} \caption{Correspondence between Gordon
    decomposition and Foldy-Wouthuysen transformation and the
    appropriate spin expressions.}
\label{diagram1}
\end{center}
\end{figure}
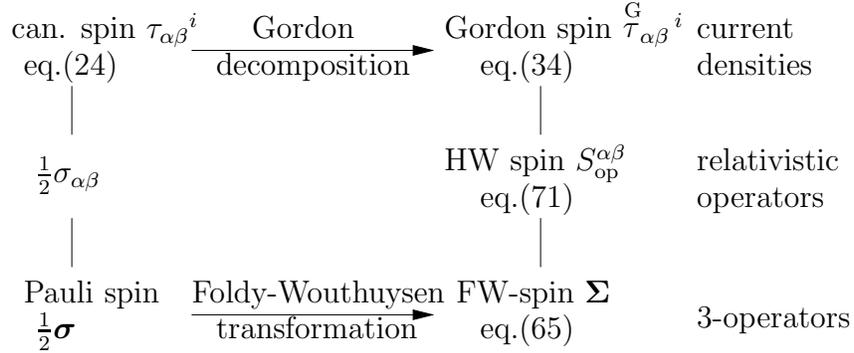

Consequently, the Gordon decomposition discussed above corresponds to
the transition from the Dirac-Pauli to the Newton-Wigner
representation of the Dirac field by means of a Foldy-Wouthuysen
transformation. In other words, the {\em Foldy-Wouthuysen
  transformation} as applied to the Dirac wave function {\em
  corresponds} to the {\em Gordon-type decomposition} in the field
theoretical picture, see Fig.\ \ref{diagram1}.

It is probably fair to say that the method of Hilgevoord-Wouthuysen to
find their conserved spin was fairly indirect.  The G\"ursey-Ryder
method is more transparent. It is an attempt to find, on an algebraic
footing, a conserved spin operator which is a representation of the
Lorentz group. Indeed, since the G\"ursey-Ryder operator satisfies the
Lorentz algebra, it generates the Lorentz group. However, this
additional requirement is not necessary for the covariance of the
expectation value of the operator. This method also contains certain
ad hoc steps as, e.g., the introduction of the $Z$-operator via
(\ref{Gue4D}).  Moreover, it turned out that, in general, it is not
conserved and thus not superior to the conventional spin
$\sigma_{\a\b}$.

On the other hand, a Gordon decomposition of the Noether currents of
momentum and spin of the Dirac field is straightforward and uniquely
determined. In this sense, the Gordon-type decomposition mentioned
yields not only the {\em gravitational moments} of the Dirac field
but, by the same token, also a decent conserved spin current --- and
this Gordon type spin current,
\begin{equation}\label{}
  \stackrel{{\rm G}}{\tau}_{\alpha\beta}{}^{\!i} = \frac{1}{4mi}
  \left[ \left(D^i\overline{\Psi}\right) \sigma_{\alpha\beta} \Psi -
    \overline{\Psi} \sigma_{\alpha\beta} \left( D^i \Psi \right)
  \right] \,,
\end{equation}
derived in a field theoretical setting, is the only one we need to
consider, since the Hilgevoord-Wouthuysen spin density can be
straightforwardly derived therefrom. By definition, spin current
densities are covariant. Thus the Gordon spin as well as its charge
are covariant and conserved, a result that shows that the Gordon spin
$\stackrel{{\rm G}}{\tau}_{\alpha\beta}{}^{\!i}$ is all we need for
a spin localization of the Dirac electron.

\subsection*{Acknowledgments}
We would like to thank to Yuri Obukhov (Moscow) for many interesting
discussions. One author (fwh) is grateful to the School for Physical
Sciences of the University of Kent at Canterbury for hospitality and,
last but not least, to the Volkswagenstiftung, Hannover, for generous
support.

\section{Appendix: Computations} 

\subsection{Proof of the identity (\ref{identity})} \label{Appendix1}
We want to prove the identity (\ref{identity})
\begin{equation} 
  \frac{1}{4} (\ol \Psi [\s^{\a\b}, \gamma^i] \Psi) - \frac{1}{4m}
  (\ol \Psi [\s^{\a\b}, \s^{ij}] D_j \Psi) = \frac{1}{2m}\ol \Psi
  (\g^i \g^\a D^\b - \g^i \g^\b D^\a) \Psi
\end{equation} 
Proof:
\begin{eqnarray}
  {\rm r.h.s} &=& \frac{1}{2m} \ol\Psi (\g^i \g^\a D^\b-\g^i \g^\b
  D^\a)\Psi \nonumber\\ &=& \frac{1}{2m}\ol \Psi (\g^i \g^\a (-im
  \g^\b \Psi + i \s^{\b j}D_j \Psi - (\a \leftrightarrow \b) )
  \nonumber\\ &=& \underbrace{\frac{i}{2} \ol\Psi (\g^i \g^\b \g^\a -
    \g^i \g^\a \g^\b)\Psi} _{A} + \underbrace{\frac{i}{2m}
    \ol\Psi(\g^i \g^\a \s^{\b j} - \g^i \g^\b \s^{\a j}) D_j \Psi}_{B}
\end{eqnarray}
\begin{eqnarray}
  A&=&\frac{i}{2}\ol\Psi(\eta^{i\b}\g^\a-\eta^{i\a}\g^\b) \Psi +
  \frac{1}{2} \ol \Psi (\s^{i\b} \g^\a - \s^{i\a} \g^\b) \Psi
  \nonumber\\ &=&\underbrace{\frac{1}{4} (\ol \Psi [\s^{\a\b},
    \gamma^i] \Psi)}_{A_1} +\underbrace{\frac{i}{2m}\ol \Psi
    (\s^{i\b}\g^\a \g^j - \s^{i\a} \g^\b \g^j)D_j \Psi}_{A_2}
\end{eqnarray}
\begin{eqnarray}
  A_2+B &=& \frac{i}{2m}\ol \Psi (\s^{i\b} \g^\a \g^j - \s^{i\a} \g^\b
  \g^j +\g^i \g^\a \s^{\b j} - \g^i \g^\b \s^{\a j} )D_j
  \Psi\nonumber\\ &=& \frac{i}{2m}\ol \Psi (\s^{i\b} \eta^{\a j} -
  \s^{i\a} \eta^{\b j} + \eta^{i \a} \s^{\b j} - \eta^{i \b} \s^{\a j}
  )D_j \Psi \nonumber\\ && + \frac{1}{2m}\ol \Psi (\s^{i\b} \s^{\a j}
  - \s^{i\a} \s^{\b j} + \s^{i \a} \s^{\b j} - \s^{i \b} \s^{\a j}
  )D_j \Psi \nonumber\\ &=& -\frac{1}{4m}\ol \Psi ( [\s^{\a\b},
  \s^{ij}] )D_j \Psi .
\end{eqnarray}
Then $A_1+A_2+B = \rm l.h.s$.

In this computation we made use of the identities 
\begin{eqnarray}
  [ \s^{\a\b}, \g^i ] = 2i ( \g^\a \eta^{\b i} - \g^\b \eta^{\a i} )
\end{eqnarray}
and
\begin{equation}\label{iden2}
  [\sigma^{ij},\sigma^{\alpha\beta}]=4i(\sigma^{i[\beta}\eta^{\alpha]j}
  -\eta^{i[\beta}\sigma^{\alpha]j}) \,.
\end{equation}

\subsection{Proof of eqn.\ (\ref{relationZ&S})}
An explicit computation (see below) of $X_{\alpha\beta}$ and
$Y_{\alpha\beta}$ yields
\begin{eqnarray}
  X_{\alpha\beta}&=&-\frac{i}{m^2} (1-\gamma_5)\, W_{\alpha\beta} +
  \frac{1}{2} \gamma_5 \sigma_{\alpha\beta}\,, \label{1}\\
  Y_{\alpha\beta}&=&-\frac{i}{m^2} (1+\gamma_5)\, W_{\alpha\beta} -
  \frac{1}{2} \gamma_5 \sigma_{\alpha\beta}\,.
\end{eqnarray}
Then $Z_{\alpha\beta}$ becomes
\begin{eqnarray}
  Z_{\alpha\beta}&:=&\frac{1}{2}(1-\gamma_5)X_{\alpha\beta} +
  \frac{1}{2}(1+\gamma_5)Y_{\alpha\beta} \nonumber\\ &=&
  \frac{-i}{m^2} (1-\gamma_5) W_{\alpha\beta}+\frac{1}{2}(1-\gamma_5)
  \frac{1}{2} \gamma_5 \sigma_{\alpha\beta}\nonumber\\ && +
  \frac{-i}{m^2}(1+\gamma_5) W_{\alpha\beta}- \frac{1}{2}(1+\gamma_5)
  \frac{1}{2} \gamma_5 \sigma_{\alpha\beta}\nonumber\\ 
  &=&\frac{-i}{m^2} 2 W_{\alpha\beta} -
  \frac{1}{2}\sigma_{\alpha\beta} = 2 S_{\alpha\beta}^{\rm op} -
  \frac{1}{2}\sigma_{\alpha\beta} \,, \label{relZ&S}
\end{eqnarray}
since $S_{\alpha\beta}^{\rm op}$ can also be written as, see
\cite{HilgevoordW}, eq.(3.7)
\begin{equation}         
  S^{\alpha\beta}_{\rm op}=\frac{1}{m^2} \varepsilon^{\alpha\beta
    \gamma\delta}P_\gamma W_\delta= \frac{-i}{m^2} W^{\a\b} \,.
\end{equation}           

\subsubsection*{Proof of (\ref{1}):}
Substitute 
\begin{equation} \label{2}
  W^{\ast \alpha\beta}=- \frac{i}{2}
  \varepsilon^{\alpha\beta\gamma\delta} \gamma_5
  W^{\ast}_{\gamma\delta} + \frac{1}{2} \sigma^{\alpha\beta} \gamma_5
  m^2
\end{equation}
into the definition of $X^{\alpha\beta}$,
\begin{eqnarray}
  X^{\alpha\beta}&:=& -\frac{i}{m^2}\left[W^{\alpha\beta} + i W^{\ast
      \alpha\beta}\right] \nonumber\\ &=&
  -\frac{i}{m^2}\left[W^{\alpha\beta} + i (- \frac{i}{2}
    \varepsilon^{\alpha\beta\gamma\delta} \gamma_5
    W^{\ast}_{\gamma\delta} + \frac{1}{2} \sigma^{\alpha\beta}
    \gamma_5 m^2) \right]\nonumber\\ &=&
  -\frac{i}{m^2}\left[W^{\alpha\beta} + i (- \frac{i}{2}
    \varepsilon^{\alpha\beta\gamma\delta} \gamma_5 \frac{1}{2}
    \varepsilon_{\gamma\delta\mu\nu} W^{\mu\nu} + \frac{1}{2}
    \sigma^{\alpha\beta} \gamma_5 m^2) \right]\nonumber\\ &=&
  -\frac{i}{m^2}\left[W^{\alpha\beta} + i (i \gamma_5 W^{\alpha\beta}
    + \frac{1}{2} \sigma^{\alpha\beta} \gamma_5 m^2)
  \right]\nonumber\\ &=& -\frac{i}{m^2} (1-\gamma_5)\, W^{\alpha\beta}
  + \frac{1}{2} \gamma_5 \sigma^{\alpha\beta}\,.
\end{eqnarray}

\subsubsection*{Proof of (\ref{2}):}
lhs.: 
\begin{eqnarray}
  W^{\ast \alpha\beta}&=&\frac{1}{2}
  \varepsilon^{\alpha\beta\gamma\delta} W_{\gamma\delta} = \frac{1}{2}
  \varepsilon^{\alpha\beta\gamma\delta} \left[-i
    \varepsilon_{\gamma\delta\mu\nu} W^\mu P^\nu \right] \nonumber\\ 
  &=&\frac{1}{2} \varepsilon^{\alpha\beta\gamma\delta} \left[-i
    \varepsilon_{\gamma\delta\mu\nu} (\frac{1}{2}
    \varepsilon^{\mu\kappa \rho\sigma} \frac{1}{2} \sigma_{\kappa\rho}
    P_\sigma) P^\nu\right]\nonumber\\ &=&\frac{1}{2}
  \varepsilon^{\alpha\beta\gamma\delta} \left[-i
    \varepsilon_{\gamma\delta\mu\nu} (\frac{1}{2} \frac{1}{2} 2i
    \sigma^{\sigma\mu} \gamma_5 P_\sigma) P^\nu\right] \,,
\end{eqnarray}
where we made use of
\begin{equation}
  \frac{i}{2} \varepsilon^{\alpha\beta\gamma\delta}
  \sigma_{\gamma\delta} =\sigma^{\alpha\beta} \gamma_5\,.
\end{equation}
1st term of rhs.:
\begin{eqnarray}
  - \frac{i}{2} \varepsilon^{\alpha\beta\gamma\delta} \gamma_5
  W^{\ast}_{\gamma\delta}&=&- \frac{i}{2}
  \varepsilon^{\alpha\beta\gamma\delta} \gamma_5
  \left[\frac{1}{2}\varepsilon_{\gamma\delta\mu\nu}
    W^{\mu\nu}\right]\nonumber\\ &=&- \frac{i}{2}
  \varepsilon^{\alpha\beta\gamma\delta} \gamma_5
  \left[\frac{1}{2}\varepsilon_{\gamma\delta\mu\nu} (-i
    \varepsilon^{\mu\nu\rho\sigma} W_{\rho} P_\sigma )
  \right]\nonumber\\ &=&- \frac{i}{2}
  \varepsilon^{\alpha\beta\gamma\delta} \gamma_5
  \left[\frac{1}{2}\varepsilon_{\gamma\delta\mu\nu} (-i
    \varepsilon^{\mu\nu\rho\sigma} (\frac{1}{2} \varepsilon_{\rho abc}
    \frac{1}{2} \sigma^{ab} P^c ) P_\sigma) \right]
 \end{eqnarray}
Here we contract the $\varepsilon$-tensors,
\begin{equation}
  \varepsilon^{\rho\sigma\mu\nu} \varepsilon_{\rho abc}=-2
  (\delta^\sigma_{[a} \delta^\mu_{b]} \delta^\nu_c+ \delta^\mu_{[a}
  \delta^\nu_{b]} \delta^\sigma_c+ \delta^\nu_{[a} \delta^\sigma_{b]}
  \delta^\mu_c)
\end{equation}
and continue
\begin{eqnarray}
  - \frac{i}{2} \varepsilon^{\alpha\beta\gamma\delta} \gamma_5
  W^{\ast}_{\gamma\delta} &=&- \frac{i}{2}
  \varepsilon^{\alpha\beta\gamma\delta} \gamma_5
  \left[\frac{1}{2}\varepsilon_{\gamma\delta\mu\nu} (-i \frac{1}{2}
    \frac{1}{2} (-2) (\sigma^{\sigma\mu} P^\nu + \sigma^{\mu\nu}
    P^\sigma + \sigma^{\nu\sigma} P^\mu ) P_\sigma) \right]\nonumber\\ 
  &=& \frac{1}{2}\varepsilon^{\alpha\beta\gamma\delta} \gamma_5
  \left[-i\varepsilon_{\gamma\delta\mu\nu} (\frac{i}{2} \frac{1}{2}(2
    \sigma^{\sigma\mu} P^\nu + \sigma^{\mu\nu} P^\sigma)P_\sigma)
  \right]\nonumber\\ &=& W^{\ast \alpha\beta} +
  \frac{1}{2}\varepsilon^{\alpha\beta\gamma\delta} \gamma_5
  \left[-i\varepsilon_{\gamma\delta\mu\nu} (\frac{i}{2} \frac{1}{2}
    \sigma^{\mu\nu} m^2) \right]\nonumber\\ &=& W^{\ast \alpha\beta} -
  \frac{1}{2} \sigma^{\alpha\beta} \gamma_5 m^2
\end{eqnarray}
This proves (\ref{2}).

\end{document}